\documentclass[structabstract]{aa}  
\usepackage{graphicx}
\usepackage{longtable}
\usepackage{txfonts}
\begin{document}
\title{A third red supergiant rich cluster in the Scutum-Crux arm}
\author{J. S. Clark\inst{1}
\and I.~Negueruela\inst{2}
\and B.~Davies\inst{3,4}
\and V.~M.~Larionov\inst{5,6}
\and B.~W.~Ritchie\inst{1,7}
\and D.~F.~Figer\inst{4}
\and M. Messineo\inst{4}
\and P.~A.~Crowther\inst{8}
\and A.~Arkharov\inst{9}}

\institute{
$^1$Department of Physics and Astronomy, The Open 
University, Walton Hall, Milton Keynes, MK7 6AA, UK\\
$^2$Departamento. de F\'{i}sica, Ingenier\'{i}a de Sistemas y
  Teor\'{i}a de la Se\~{n}al, Universidad de Alicante, Apdo. 99, E03080
  Alicante, Spain\\
$^3$School of Physics \& Astronomy, University of Leeds, Woodhouse Lane, Leeds, LS2 9JT, UK\\
$^4$Chester F. Carlson Centre for Imaging Science, Rochester Institute of Technology, 54 Lomb Memorial Drive, Rochester NY 14623, USA\\ 
$^5$ Astronomical Institute of St. Petersburg University, Petrodvorets, Universitetsky pr. 28, 198504 St. Petersburg, Russia\\
$^6$ Isaac Newton Institute of Chile, St.Petersburg branch \\
$^7$ IBM United Kingdom Laboratories, Hursley Park, Winchester, Hampshire, SO21 2JN, UK.\\
$^8$ Department of Physics \& Astronomy, University of Sheffield, Sheffield, S3 7RH, UK\\
$^9$ Pulkova Astronomical Observatory, 196140 St. Petersburg, Russia
}

\abstract{}{We aim to characterise the properties of a third massive, red supergiant
  dominated galactic cluster.}{To accomplish this we utilised a combination of near/mid-
  IR photometry and spectroscopy  
to identify and classify the properties of cluster members, and
statistical arguments to determine the mass of the cluster.} 
{We found a total of 16 strong candidates for cluster membership, for
  which formal classification of a subset yields  spectral types from
  K3--M4\,Ia and luminosities 
between $\log(L/L_{\odot})\sim4.5$--4.8 for an adopted distance of $6\pm1\:$kpc. For an age in the range of 16--20\,Myr, the implied mass is
 2--$4\times10^4\,$M$_{\odot}$, making it one of the most massive young clusters
 in the Galaxy. This discovery supports the hypothesis that a
 significant burst of star formation  
occurred at the base of Scutum-Crux arm between 10--20\,Myr ago,
yielding a stellar complex comprising at least $\sim10^5$M$_{\odot}$  of stars (noting that since the
cluster identification criteria rely on the presence of RSGs, we suspect that the true stellar yield 
will be significantly higher). We  highlight the apparent absence  of X-ray binaries 
within the star formation complex and finally, given the  physical association  of at least 
two pulsars with this region, discuss the implications  of this  finding 
for stellar evolution and the  production and properties of neutron stars.}{}


\maketitle

\section{Introduction}

The vigorous  star formation that characterises starburst galaxies
 results in the production of extended complexes of young massive
 stellar clusters, which  
span hundreds of parsecs but appear to have formed over a limited
time-frame ($\leq20\:$Myr; Bastian et al. \cite{nate05}). With masses  
$>10^4\:$M$_{\odot}$, analogues of such constituent clusters had been
thought to be absent from our own  
Galaxy. However near-IR  observations revealed that the Galactic Centre
hosts 3 such young  massive  
 clusters (Figer et al. \cite{figer99}, \cite{figer02}, \cite{figer04}),
 while detailed study of  \object{Westerlund~1} suggests a mass of the
 order of $10^5\:$M$_{\odot}$  
(Clark et al. \cite{clark05}). Such discoveries  raise the exciting
prospect of {\em directly} determining such 
 fundamental properties  as their (Initial) Mass Function; currently 
impossible for unresolved extragalactic examples.

Furthermore, their presence in the Galaxy 
permits the detailed investigation of massive stellar evolution, since
their high mass 
yields significant  numbers of  rare spectral types in a co-eval
setting of uniform metallicity.  With ages  
$<5\:$Myr, \object{Westerlund~1} and the Galactic Centre clusters
provide valuable insights into the properties and 
evolutionary pathways of massive ($>40\:$M$_{\odot}$) stars. Recently,
studies  by Figer et al.  
(\cite{figer06}; F06) and Davies et al. (\cite{davies07},
\cite{davies08}; D07 \& D08 respectively) have revealed two  
further massive clusters dominated  by red supergiants (RSGs) at the
base of Scutum-Crux arm --  RSGC1 ($12\pm 
2\:$Myr; $M_{{\rm initial}}=3\pm1\times10^4\:$M$_{\odot}$)  and RSGC2
($17\pm3\:$Myr; $M_{{\rm
    initial}}=4\pm1\times10^4\:$M$_{\odot}$). Collectively, both clusters
sample a somewhat lower range of stellar masses, hosting  40 RSGs with  
$M_{{\rm initial}}\sim14-20\:$M$_{\odot}$; of particular interest since such stars are thought to be Type II SNe
progenitors (Smartt et al. \cite{smartt}). 

In this paper we report the discovery of a third massive, RSG
dominated cluster, RSGC3,  also located at the base of the Scutum-Crux arm.  
Identified visually in GLIMPSE/Spitzer mid-IR images (Benjamin et al. 
\cite{benjamin}) as a concentration
of bright stellar sources at $\sim l=29\fd2, b=-0\fd2$,  
we  utilised near-IR photometry to identify  potential cluster
members, a subset of which were subsequently observed
spectroscopically to  
provide a firm classification. Finally, a synthesis of   these data
were used to constrain the bulk properties of the cluster and
individual stars  
within it,  enabling a comparison to RSGC1 \& 2 and a
characterisation of the star forming environment they delineate.

\section{The RSG candidate sample}

\begin{figure*}
\centering
   \resizebox{\columnwidth}{!}{\includegraphics[width=12cm]{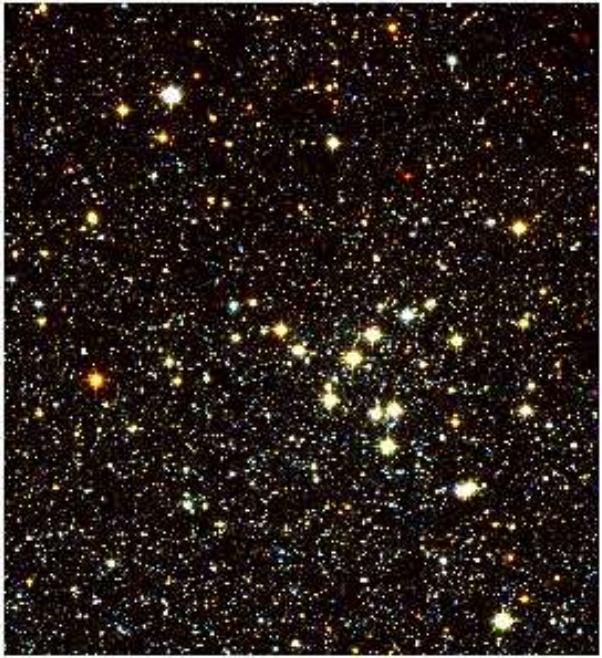}}
\caption{Near-IR $JHK$-band colour composite  of the central
  $\sim7\farcm5\times8\farcm5$ of RSGC3 ($\sim13.1\times 14.9$\,pc at a distance of 6\,kpc),
 constructed from  UKIDSS data (Lawrence et al. \cite{lawrence})
with artifacts due to saturation artificially
removed. Note the lack of a clearly defined stellar  
overdensity of unevolved cluster members with respect to the field.}
\end{figure*}

\begin{figure*}
\centering
   \resizebox{\columnwidth}{!}{\includegraphics{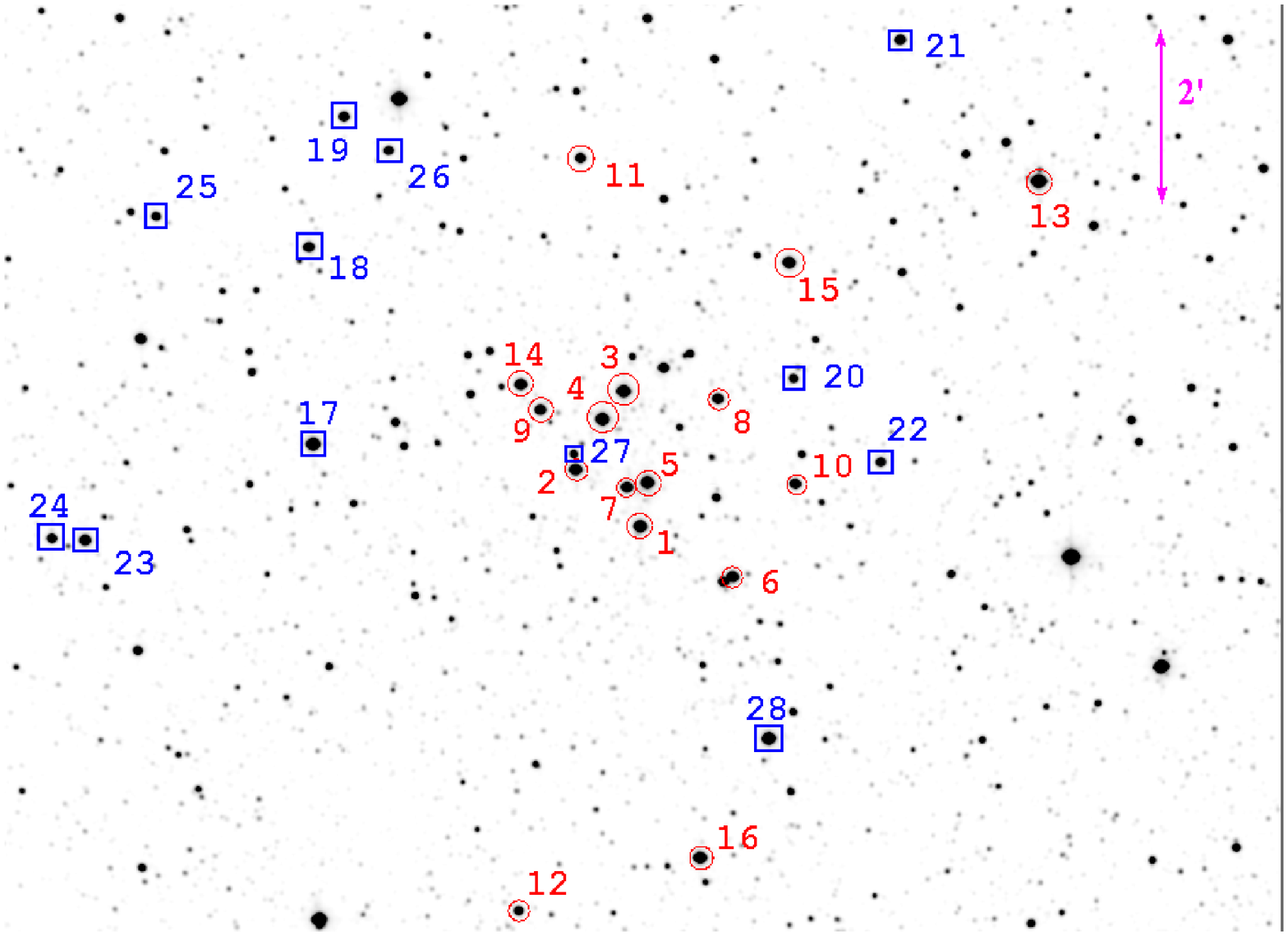}}
\caption{Finding chart for RSGC3, with the stars listed in Table~1
  indicated. The finder comprises a $K_{\rm s}$-band image from 2MASS with a 
$14\arcmin\times10\arcmin$ field of view centred on the cluster ($\sim24.5\times17.5$\,pc for a distance of 6\,kpc).
Note that this represents a larger field of view than the near-IR image presented in Fig~1.
The circles represent stars we consider to be  {\em bona fide}  cluster members,
 while the squares indicate the remaining stars discussed in the text, 
a subset of which (S17-22) are also likely cluster members (see Sect. 3 for details).}
\end{figure*}

As can be seen from the near IR images of RSGC1--3 (F06, D07 \& Fig.~1
\&~2) it is extremely difficult to determine  
 a physical extent  for such (putative) clusters since, with the exception of the
 RSGs, no other cluster population is readily visible 
 as an overdensity with respect to the stellar field
 population. If kinematic information is available, it is possible to identify a 
co-moving,  physical association of RSGs, in order to discriminate between
 cluster and field stars (e.g. D07). However, the 
spectroscopic data presented here are of insufficient resolution to
extract the radial velocity of cluster members, while,  
unlike RSGC1 \& 2 (F06; D07), we find no maser emission from any
cluster members which would also provide kinematic  
constraints (Verheyen et al. in prep.);  thus  we are forced to utilise
   photometric data to construct  a list of candidate cluster members.

 Based on the spatial concentration of bright red stars, we start by
taking 2MASS photometry for stars within $r\leq3\arcmin$ of the position
of Star 1 (RA: 18h 45m 23.60s, DEC:
$-03\degr\:24\arcmin\:13\farcs9$), selecting only
stars with quality flags "AAA" and error $\Delta K_{{\rm S}}\leq0.05$. The
ten bright stars defining the spatial concentration form a
well-separated group
in the $(J-K_{{\rm S}})/K_{{\rm S}}$ diagram (Fig.~3), around $(J-K_{{\rm
S}})\approx3.0$. This grouping is also present in the $(H-K_{{\rm S}})/K_{{\rm S}}$ diagram, centred around
$(H-K_{{\rm S}})=0.95$. We then calculate the reddening-free parameter $Q_{{\rm
IR}}=(J-H)-1.8\times(H-K_{{\rm S}})$. Early-type stars have $Q_{{\rm
IR}}\approx 0.0$, while most bright field stars have $Q_{{\rm IR}}\approx
0.4-0.5$, corresponding to red  giants (Indebetouw et al. \cite{ind}, 
Negueruela \& Schurch \cite{negueruela}). All ten stars form a
clearly separated grouping in this diagram, with values $0.2-0.4$,
typical of supergiants. There is one more star in this clump,
S14, which has redder $(J-K_{{\rm S}})$ and $(H-K_{{\rm S}})$. The
only other star of comparable $K_{{\rm S}}$ in the field, S28, has
$Q_{{\rm IR}}=0.08$, typical of an early-type star.
Considering the large number of bright stars in the field and the
spatial extent of other starburst clusters in the area (e.g. F06,
D07), we extended 
the search to $r\leq7\arcmin$. The group in the  $(J-K_{{\rm
S}})/K_{{\rm S}}$ and  $(H-K_{{\rm S}})/K_{{\rm S}}$ diagrams, 
which we consider to comprise prime cluster members  now includes S11,
S12, S13 and S15 (Table~1).

A number of objects -- S14 (found in the cluster core), S15--16 and S18--22 --
have $Q_{{\rm IR}}$ similar to the above, but with redder  -- 
$(J-K_{{\rm S}})\approx3.7$ -- colours; the  separation between these
stars and the main plume of red giant stars in terms of $Q_{{\rm
IR}}$ is not as well defined as that for the prime cluster candidates.  
We identify these as likely cluster members, with the difference in
colours potentially due   
to excess reddening with respect to the core members (noting that
significant differential reddening  is also observed for RSGC2; D07).
Finally, for completeness, with the inclusion of S23 \& S27, these
stars form a well defined group in the $(J-K_{{\rm S}})/K_{{\rm S}}$
diagram. However, only S14, S15, S16 \& S18  are grouped in  
the $(H-K_{{\rm S}})/K_{{\rm S}}$, while S19, S20, S22, S24, S25, S26
\& S27  form a second, distinct 
 group  in $(H-K_{{\rm S}})$. Therefore,  given their magnitudes and red 
 colours, we identify S23--27 as
 potential objects of  interest, but as with our second group, they require  
spectroscopic follow up to ascertain their nature and relationship to
RSGC3 (Table~1).   

To summarise, based solely on their near-IR properties we identify a
core group of 15 prime candidate 
cluster members, a second group of 7 likely cluster members, and a
final group of 5 bright red stars 
 that deserve investigation within $r\leq7\arcmin$ of the nominal
 cluster core (Table~1).

\begin{figure}
\includegraphics[width=5.7cm]{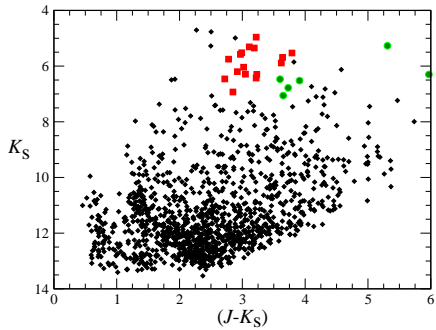}
\caption{Colour magnitude plot for stars within $7\arcmin$ of
  RSGC3. The 16 cluster members identified in Sect.~3 are  
indicated by the red squares, the remaining 6 likely cluster members
(Sect.~2 \& Table~1) are plotted as green circles.  
 Note that the two outlying stars with $(J-K_{{\rm S}})>5.0$ are S17
 and S21, discussed in  Sect.~3. Finally, of the two stars which are
 also located with the (likely) cluster members and not marked as
 such, one is a blend and hence we exclude it from further analysis,
 while the second is S23, which has a discrepant  
$(H-K_{{\rm S}})$ index and $Q_{{\rm IR}}=-0.4$, suggesting that it
may be an emission line star rather than a RSG.} 
\end{figure}

\section{Spectroscopic results and analysis}

Based on their photometric properties, initial low resolution ($R
\sim270$) observations of 17 stars were made with the 
IR imaging camera SWIRCAM+HK grism, mounted on the AZT-24 1.1m telescope at Campo Imperatore on 2006 September 3 \& 4. 
Subsequently, higher resolution  observations of 8 targets, made in
the flexible observing  mode, were obtained  with the $1-5\:\mu$m imaging spectrometer UIST,
mounted on the  United  
Kingdom Infra-Red Telescope on 2007 June 8 \& 21 (Program ID
U/07A/15). The Long K grism was used with the 4-pixel slit, giving wavelength coverage from 2.20-$2.51\:\mu$m with a 
resolution $R \sim 1900$.  Data reduction was accomplished via the methodology described in Clark et
al. (\cite{clarkLBV}), and the spectra are presented 
in Fig.~2.

Of the 17 low resolution spectra, 16 show deep  CO bandhead
absorption,  characteristic of late type stars (Fig.~4). Of  
these, 14 are photometrically defined core cluster members; one, S16,
a likely member, and the final star, S28, appears to be a  
foreground object based on its  near-IR colours (Table 1). Following the
methodology of F06 and D07, it is possible to use the  
strength of the CO bandheads to provide a spectral and luminosity
classification for the stars. However this requires  
a robust determination of the stellar  continuum, which proved
impossible for the low resolution spectra, and  
consequently was only attempted for the subset of 8 stars for which
medium resolution data were available.  

We find  all 8 stars -- S2-5 \& S7-10 --  to be supergiants, with spectral types ranging from  K5--M4\,Ia; the resultant 
temperatures (and associated errors) are summarised in Table~2. Given
the equivalence of the low resolution spectra  
of these stars with those of S1, S6, S11 \& S13--16, we conclude that
these stars are likewise RSGs. Thus these 
 results provide strong support for the identification of S1--15 as
 {\em bona fide} cluster members based on both  
spectroscopic and photometric criteria, with S16 possibly a more
heavily reddened cluster member. For the remainder of  
the paper we therefore count these sixteen stars as cluster members; in a future 
paper we will use high resolution spectroscopy to 
confirm such a physical association (Davies et
 al. in prep.). 

The {\em current} lack of kinematic data  precludes the  determination of the
 cluster distance via comparison to the Galactic rotation curve, and hence the  
luminosity, age and initial mass of the cluster members  (since RSGs
span a wide range of  luminosities   
($\log(L_{{\rm bol}}/L_{\odot})\sim4.0$--5.8; Meynet \& Maeder
\cite{meynet}). Nevertheless, we find a mean value of $A_{K_{S}} \sim1.5$,  
from which we may infer  $A_V \sim13.0$ and hence, assuming a canonical
1.8\,mag. extinction per kpc,  
 an upper limit to the distance to RSGC3 of $\sim$7.2\,kpc (Rieke \& Lebofsky \cite{rieke}, Egan et al. 
\cite{egan02}).  Such a value is
 entirely consistent with a location of RSGC3 at a similar distance to  
RSGC1 \& 2 at the end of the Galactic Bar (6.60$\pm$0.89\,kpc and 5.83$^{1.91}_{-0.76}$\,kpc respectively,
 D08); we thus adopt a distance of
$6\pm1\:$kpc for the remainder of this work. 
At such a distance, utilising the temperature/spectral type calibration and resultant bolometric corrections
of Levesque et al. (\cite{levesque}), we find  $\log(L_{{\rm bol}}/L_{\odot})\sim4.5$--4.8 for S2--5 \& 7--10 and  corresponding ages and initial masses  of  
$16--20\:$Myr and $\sim10$--$13\:$M$_{\odot}$  (Fig.~5).  

Finally, the  observed range of the dereddened [8]--[12] colour index -- 
{\it MSX} $(A-C)\sim0.85$--1.44 (Table 1) -- for the 7 (candidate)
cluster members for which it may be determined is directly comparable
to that found by  D07 for RSGC2. 
Sampling the broad silicate emission feature, this provides a measure
of the mass loss rate via the dust content of the circumstellar  
environment. It is therefore of interest that both of the stars with
discrepant ($J-K_{{\rm S}}$)  colours show  excesses (S17 \& S12;
Fig 3),  suggesting that a build-up of circumstellar material due to
enhanced mass-loss affects their near-IR properties, such as is observed
in RSGC2-49 (D07).

\begin{figure} 
\includegraphics[width=8cm]{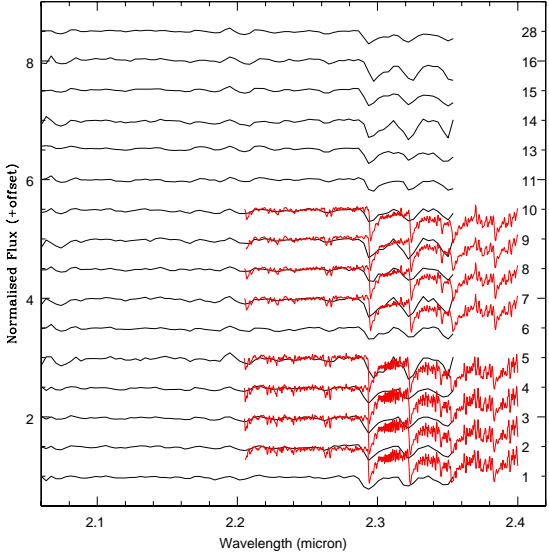}
\caption{Montage of low (Campo Imperatore; black) and medium resolution (UKIRT; red) spectra of selected
photometric targets, revealing  the prominent CO bandhead absorption.}
\end{figure}

\begin{figure}
\includegraphics[width=8cm]{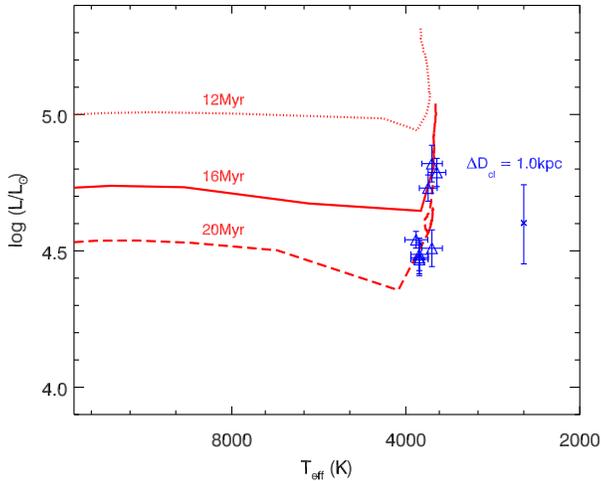}
\caption{H-R diagram showing the locations of the 8 RSGs for which accurate spectral classification was possible assuming 
a distance to the cluster of $D_{\rm cl} \sim6\:$kpc. 12, 16 \& 20\,Myr isochrones
from the rotating models of Meynet \& Maeder (\cite{meynet}) have been overplotted. Errors on the data points do not include the uncertainty
 in the cluster distance ($\Delta D_{{\rm cl}}$);  representative error-bars assuming an uncertainty of $\pm1\:$kpc are indicated to the right of the 
figure.}
\end{figure}

\begin{table*}
\begin{center}
\caption{Summary of RSG candidates and their properties. Top panel:
  the core group of 15 {\em photometrically} selected stars 
regarded as prime cluster candidates; second  panel: likely cluster
members;  third panel: stars of interest, as defined in Sect.~2 and fourth panel: likely foreground
RSG identified by spectroscopy and included for completeness.
Note that based on spectroscopy, we also consider it likely that star
S16 is a cluster member and treat it as such in the text. Co-ordinates
and near-IR magnitudes are from 2MASS, with mid-IR ($\sim5$--25$\mu$m)
magnitudes from the Galactic plane surveys of GLIMPSE/{\it Spitzer}
(Benjamin et al. \cite{benjamin}) and the {\it Midcourse Source
  Experiment (MSX)}  (Egan et al. \cite{egan}). We also 
provide the dereddened {\it MSX} $(A-C)$ colour (adopting the
prescription of Messineo et al. \cite{maria}), which is 
 a diagnostic of emission from circumstellar dust (and hence proxy for
 mass loss) and the spectral type  of the stars, where available (Sect.~3).}
\begin{tabular}{lccccccccccccc}
\hline
\hline
ID & \multicolumn{2}{c}{Co-ordinates (J2000)}  & & 2MASS &  & GLIMPSE &  {\it
  Spitzer} & & & {\it MSX} & & & Spec. \\
   &      RA     &    Decln.  &   $J$ & $H$    & $K_{\rm s}$ &   5.8$\mu$m  & 8.0$\mu$m  &  A   &   C  &   D   &  E & (A-C)  &  Type \\
\hline
S1 & 18 45 23.60 &-03 24 13.9 & 8.55& 6.54 &5.58  &  4.73 & 4.50 &  -   &   -  &   -   &  - &   -    &     RSG \\
S2 & 18 45 26.54 &-03 23 35.3 & 8.53& 6.62 &5.75  &  4.92 & 4.63 & 4.30 & 3.39 &   -   &  - &   -    &     M3\,Ia \\
S3 & 18 45 24.34 &-03 22 42.1 & 8.54& 6.43 &5.35  &  4.19 & 4.14 &  -   &   -  &   -   &  - &   -    &     M4\,Ia \\
S4 & 18 45 25.31 &-03 23 01.1 & 8.42& 6.39 &5.31  &  4.08 & 4.04 & 3.97 & 3.12 &  3.21 &  - &  0.85  &     M3\,Ia \\
S5 & 18 45 23.26 &-03 23 44.1 & 8.51& 6.52 &5.52  &  4.45 & 4.20 & 3.78 & 2.82 &  3.15 &  - &  0.96  &     M2\,Ia \\
S6 & 18 45 19.39 &-03 24 48.3 & 9.06& 6.97 &6.04  &  5.24 & 5.20 & 4.71 &   -  &   -   &  - &   -    &     RSG \\
S7 & 18 45 24.18 &-03 23 47.3 & 9.12& 7.10 &6.20  &  5.22 & 4.84 &  -   &   -  &   -   &  - &   -    &     M0\,Ia \\
S8 & 18 45 20.06 &-03 22 47.1 & 9.53& 7.29 &6.30  &  5.46 & 5.43 & 5.60 &   -  &   -   &  - &   -    &     K5\,Ia \\
S9 & 18 45 28.13 &-03 22 54.6 & 9.34& 7.26 &6.29  &  5.49 & 5.44 &  -   &   -  &   -   &  - &   -    &     M0\,Ia \\
S10 & 18 45 16.56 &-03 23 45.1 & 9.65& 7.43 &6.43  &  5.57 & 5.52 & 5.45 &   -  &   -   &  - &   -    &    M0\,Ia \\
S11 & 18 45 26.31 &-03 20 03.3 & 9.18& 7.27 &6.46  &  5.77 & 5.78 & 5.87 &   -  &   -   &  - &   -    &     RSG \\
S12 & 18 45 29.12 &-03 28 35.6 & 9.78& 7.86 &6.93  &  6.57 & 6.26 & 6.23 &   -  &   -   &  - &   -    &      -  \\
S13 & 18 45 05.51 &-03 20 19.6 & 8.18& 5.97 &4.96  &    -  &  -   & 3.59 & 2.43 &  2.45 &  - &  1.16  &     RSG \\
S14 & 18 45 29.02 &-03 22 37.4 & 9.51& 7.08 &5.89  &  4.79 & 4.61 & 4.43 &   -  &   -   &  - &   -    &    RSG \\
S15 & 18 45 16.84 &-03 21 14.6 & 9.33& 6.92 &5.69  &  4.12 &   -  & 3.26 & 1.85 & 1.66  & 0.92 &1.41  &    RSG \\
\hline
S16 & 18 45 20.87 &-03 27 59.6 & 9.32& 6.78 &5.53  &  4.34 & 4.15 & 3.79 & 2.93 & 2.97  & 2.22 &0.86  &    RSG \\
S17 & 18 45 38.45 &-03 23 18.1 &10.58& 7.07 &5.27  &    -  &   -  & 2.09 & 0.65 & 0.45  &-0.121 & 1.44&     - \\ 
S18 & 18 45 38.65 &-03 21 03.9 &10.07& 7.62 &6.47  &  5.50 & 5.44 & 5.36 &   -  &   -   &  -    &-    &     - \\
S19 & 18 45 37.05 &-03 19 35.1 &10.43& 7.81 &6.52  &  5.44 & 5.40 & 5.42 &   -  &   -   &  -    &-    &     - \\
S20 & 18 45 16.64 &-03 22 33.2 &10.71& 8.28 &7.06  &  5.98 & 5.85 & 5.79 &   -  &   -   &  -    &-    &     - \\
S21 & 18 45 11.81 &-03 18 43.1 &12.27& 8.30 &6.30  &  4.50 & 4.32 & 4.27 & 3.31 & 3.17  &  -    &0.96 &     - \\
S22 & 18 45 12.69 &-03 23 30.7 &10.51& 8.04 &6.78  &  5.49 & 5.41 & 5.36 &   -  &  -    &  -    &-    &     - \\
\hline
S23 & 18 45 48.82 &-03 24 23.1 & 9.67& 7.36 &5.85  &    -  &  -   & 2.92 & 2.00 & 1.84  & 1.27  &0.92 &     - \\
S24 & 18 45 50.34 &-03 24 21.8 &10.84& 8.08 &6.74  &    -  &  -   & -    & -    & -     & -     &-    &     - \\
S25 & 18 45 45.61 &-03 20 42.8 &11.27& 8.43 &7.10  &  5.88 & 5.96 &  -   & -    & -     & -     &-    &     - \\
S26 & 18 45 35.02 &-03 19 58.2 &11.43& 8.43 &6.96  &  5.42 & 5.21 &  -   &  -   &   -   & -     &  -  &     - \\
S27 & 18 45 26.61 &-03 23 24.7 &11.26& 8.78 &7.46  &  6.01 & 5.81 &  -   &-     &-      & -     &-    &     - \\
\hline
S28 & 18 45 17.75 &-03 26 38.4 & 7.78& 6.14 &5.38  &  4.42 & 4.17 & 3.18 & 3.12 & -     & -     &1.19 & RSG   \\
\hline
\end{tabular}
\end{center}
\end{table*}

 \begin{table}
\caption{Summary of the stellar properties of the 8 RSGs for which accurate spectral classification was possible,
assuming a cluster distance of 6\,kpc.}
  \begin{tabular}{lccccc}
        \hline \hline
        ID\# & $T_{{\rm eff}}$ & Spec & $A_{K}$ & $M_{K}$ &
        $\log(L_{{\rm bol}}/L_{\odot})$ \\
             & (K) & Type &       &      &     \\
        \hline
 2 & 3605$\pm$147 & M3 & 1.20$\pm$0.12 &  -9.34$^{+0.39}_{-0.34}$ &   4.51$^{+0.14}_{-0.15}$ \\
 3 & 3535$\pm$125 & M4 & 1.47$\pm$0.08 & -10.09$^{+0.37}_{-0.33}$ &   4.79$^{+0.13}_{-0.15}$ \\
 4 & 3605$\pm$147 & M3 & 1.44$\pm$0.12 & -10.12$^{+0.39}_{-0.34}$ &   4.82$^{+0.14}_{-0.15}$ \\
 5 & 3660$\pm$130 & M2 & 1.43$\pm$0.07 &  -9.85$^{+0.37}_{-0.32}$ &   4.73$^{+0.13}_{-0.15}$ \\
 7 & 3790$\pm$130 & M0 & 1.40$\pm$0.11 &  -9.11$^{+0.38}_{-0.33}$ &   4.48$^{+0.13}_{-0.15}$ \\
 8 & 3840$\pm$175 & K5 & 1.60$\pm$0.04 &  -9.23$^{+0.36}_{-0.31}$ &   4.54$^{+0.12}_{-0.14}$ \\
 9 & 3790$\pm$130 & M0 & 1.53$\pm$0.11 &  -9.14$^{+0.38}_{-0.33}$ &   4.49$^{+0.13}_{-0.15}$ \\
10 & 3790$\pm$130 & M0 & 1.60$\pm$0.11 &  -9.09$^{+0.38}_{-0.33}$ &   4.47$^{+0.13}_{-0.15}$ \\
   \hline
   \end{tabular}
\end{table}

Following the methodology pioneered by F06, we utilise Monte Carlo
simulations, employing  rotating stellar models (D07) to estimate the
initial mass of the cluster from the number of RSGs currently
present. {\em For ages of 16(20)\,Myr this yields masses of
  $2(4)\times10^4\:$M$_{\odot}$}.  If our second tier of RSG
candidates are confirmed as cluster members, the mass estimate would
increase by $\sim$30\%, noting that this would  not be altered by  
increasing the distance to a maximum of 7\,kpc (as implied by the
reddening). Finally, since stellar evolutionary codes predict 
a spread in intrinsic RSG  luminosities  for even co-eval clusters, the observed range of
log(L$_{bol}$/L$_{\odot}$)$\sim$4.5-4.8 for cluster members should not be interpreted as
implying non-coevality.

\section{Discussion and concluding remarks}

With an age of 16--20\,Myr and a total mass of
2--4$\times10^4\:$M$_{\odot}$,  RSCG3 appears to be a close counterpart
to RSGC1 \&~2 (D08), while the properties of the constituent stars in 
terms of spectral types, luminosities and circumstellar environments are 
also directly comparable to the members of those clusters. As such RSGC3 
belongs to an increasing population of hitherto unsuspected young massive clusters 
within the Galaxy.  With ages ranging from 2-3\,Myr for the Arches to 16-20\,Myr for 
RSGC2 \& 3 they also provide a fertile testbed for constraining the lifecycle of  
stars of $\sim14\:$M$_{\odot}$ and above. 

Additionally, the proximity of RSGC3  to both RSGC1 \&~2 --  projected distances of $\sim400\:$pc \& $\sim300\:$pc
respectively for d=6~kpc --  at the base of Scutum-Crux 
arm (l$\sim25-29^{\rm o}$) provides strong support for the hypothesis that this region has
been subject to a recent burst of star formation (Garz\'on et
al. \cite{garzon}, D07), yielding an extended stellar cluster complex such as those observed 
in external galaxies such as M51 (Bastian et al. \cite{nate05}).  If correct, the   
`starburst' has yielded a total of $>9\times10^4\:$M$_{\odot}$ of stars
just considering the 3 clusters. However, D07 report
 the presence of  an additional  population of RSGs in the vicinity of RSGC2, while 
 Garz\'on et al. (\cite{garzon}) and L\'opez-Corredoira et al. 
(\cite{lopez}) also report a significant `diffuse' field population
of cool supergiants within the region delineated by RSGC1-3, suggesting that the true total may
be significantly higher. 

Consideration of the near-IR images of RSGC1--3 emphasises this
possibility; the clusters are only identifiable as such due to their
significant RSG populations, as no other  overdensity of (less evolved) stars
 is apparent.   However, such stars  
are intrinsically  short lived and  consequently rare, and so only signpost the
location of clusters for which the (unevolved) stellar  population is
unresolvable against the field for a  
narrow range of cluster masses and ages. For instance, for clusters
with masses of the order of  $10^{3}\:$M$_{\odot}$ (so comparable to the Orion cluster) one would
only  expect the presence of $\sim$1-3 RSGs at any given epoch on statistical grounds,
suggesting that such clusters would be difficult to identify  in a near-IR
survey (a problem even  afflicting more massive clusters for ages  $<10\:$Myr; D07).

Likewise, for clusters with ages of over 20\,Myr, while one would expect a large number of red evolved 
stars to be present, their lower intrinsic luminosity will make them more difficult to discriminate against the 
field in any  search for cluster candidates.  Furthermore, the rapid dissolution of clusters 
due to ejection of the intercluster medium by stellar winds and SNe (Goodwin \& Bastian
\cite{goodwin}) exacerbates this problem, reducing the spatial density of any RSGs within clusters. An additional 
result of this process is that the velocity dispersion of cluster members will increase, in turn magnifying  the 
uncertainty in the identification of a physical association via kinematic means.  

Therefore, while the presence of RSGC1--3 points to an episode of
enhanced star formation 10--20\,Myr ago, from the methodology employed in this 
work  it is difficult to determine whether there was significant activity 
before this date or indeed what the total stellar yield of this 'starburst' was. An analogous
argument applies to determining the star  formation history for ages $<10\:$Myr, although the lack of Giant
H\,{\sc ii} regions within this putative complex (Conti \& Crowther
\cite{conti}) implies that no massive clusters are currently forming.

We caution that these limitations will manifest themselves in  any other near-IR searches for stellar clusters 
in regions of the  disc with a high (projected) stellar density, likely leading  to significant incompleteness in any
 Galactic cluster mass function determined via such a methodology.

\subsection{Association with X-ray binaries and post SNe compact objects}

Given a potential SNe rate of one every 40--80\,kyr (F06), and the
likely association of two pulsars with RSGC1 (F06, Gotthelf \& Halpern
\cite{gotthelf}) we examined the  
catalogues of Liu et al. (\cite{liu}) and Bird et al. (\cite{bird}) to
search for any relativistic sources associated with RSGC3, but found
none. Motivated by the hypothesis  
that RSGC1--3 delineate a star forming complex, we extended this
search to this region ($l=25-30\degr$, $b=\pm1\degr$), but again found
no accreting X-ray binaries within it for a  
distance of  $6\pm1\:$kpc. Given the increasing evidence for a high
binary fraction amongst massive stars (e.g., Clark et
al. \cite{clark08}), the  absence of any Be/X-ray binaries --  
systems consisting of a B0--3 V-IIIe primary and a neutron star accretor -- is surprising, since one would expect them to 
be active at such an epoch (10--20\,Myr; e.g., Portegies Zwart \&
Verbunt \cite{PZ}). Nevertheless, one might suppose that a combination
of their transient nature plus a SNe kick  
sufficient to either disrupt or rapidly eject a surviving binary from
the complex may explain their lack of detection.  
 
However, we note with interest the location of the Anomalous X-ray
Pulsar (AXP) \object{AX J1841.3$-$0455}; $l=27\fd39$,  $b=-0\fd006$.  
Durant \& van Kerkwijk (\cite{durant}) estimate a lower limit of
$>5\:$kpc to \object{AX J1841.3$-$0455}, while  
Vasisht \& Gotthelf (\cite{vasisht}) provide an upper limit of 7\,kpc
from its association with the SNR \object{Kes 73}. Taken together
they  raise the possibility that it could be physically associated
with the putative star formation complex, with \object{AX J1841.3$-$0455}
located $\sim$equidistantly between RSGC2 \& 3, and directly within the
region identified by  Garz\'on et al. (\cite{garzon}) and L\'opez-Corredoira
 et al. (\cite{lopez}) as showing a significant overdensity of 'field' RSGs.
 The expected rate for SNe for such a complex would be fully  
consistent with the relative youth expected of magnetars (Thompson et
al. \cite{thompson}), while \object{SGR 1900+14} demonstrates that
their  progenitors can have masses as low as $\sim15\:$M$_{\odot}$ (Clark
et al. \cite{clark08}, Davies et al. in prep.), consistent with the
current RSG population at the base of the Scutum Crux arm.  

Definitively associating \object{AX   J1841.3$-$0455} with the same burst of star 
formation that yielded RSG1--3 would imply a progenitor mass of $<20\:$M$_{\odot}$, and 
hence provide additional evidence that the hypothesis that high-mass stars are required to produce          
magnetars is incorrect. Moreover,  consideration of 
\object{SGR 1900+14}, \object{AX J1838.0-0655} (the young pulsar associated with RSGC1)
and {\em potentially} \object{AX   J1841.3$-$0455}  suggests  that despite having 
progenitors of comparable mass ($\sim15-18\:$M$_{\odot}$; Clark et al. \cite{clark08}, D08, Davies et al. in prep.), 
the surface magnetic fields of the resultant  neutron stars can differ by over two orders of magnitude
(Gotthelf \& Halpern \cite{gotthelf}, Kouveliotou et al. \cite{kou}, 
Vasisht \& Gotthelf  \cite{vasisht}), presumably reflecting   differences 
in the properties of their progenitors  {\em other} than, or being directly dependant on, 
stellar mass (such as magnetic field or rotational velocity).

\begin{acknowledgements}

JSC acknowledges support from an RCUK fellowship, and thanks Sophie
Allen for assistance in the preparation of Fig.~1, and Mike \& Tessa Allen for their kind hospitality
during the production of this paper.
This research is partially supported by the Spanish Ministerio de
Ciencia e Innovaci\'on (MICINN) under
grants AYA2008-06166-C03-03 and CSD2006-70.
AZT-24 observations are made within     
an agreement between Pulkovo, Rome and Teramo observatories
D.F. acknowledges support from NASA under award NNG 05-GC37G,
through the Long-Term Space Astrophysics program and from NYSTAR
under a Faculty Development Program grant. The UKIDSS project is defined in
Lawrence et al. (\cite{lawrence}). UKIDSS uses the UKIRT Wide Field Camera (Casali et al. \cite{casali}). The photometric 
system and calibration are described in Hewett et al. (\cite{hewett}) and Hodgkin et al. (\cite{hodgkin}), with pipeline 
processing and archiving described in Irwin et al. (in prep.) and Hambly et al. (\cite{hambly}).  This paper makes use of data products from the Two Micron All Sky Survey, which is a joint project of the University of Massachusetts and the Infrared Processing and Analysis Center/California Institute of Technology, funded by the National Aeronautics and Space Administration and the National 
Science Foundation.

\end{acknowledgements}

{}
\end{document}